\documentclass[twocolumn,showpacs,preprintnumbers,amsmath,amssymb]{revtex4} 
\usepackage{bm}
\input{psfig.sty} 
\begin{document} 
\preprint{Physical Review {\bf 66}, 016125 (2002)}

\title{Scaling exponents and clustering coefficients of a 
growing random network}
\author{Haijun Zhou}
\affiliation{Max-Planck-Institute of Colloids and Interfaces,
Potsdam 14424, Germany}

\date{\today}

\begin{abstract}
The statistical property of a growing scale-free network is studied
based on an earlier model proposed by Krapivsky, Rodgers, and
Redner [Phys.\ Rev.\ Lett.\ {\bf 86}, 5401 (2001)], with the additional
constraints of forbidden of self-connection and multiple links of the
same direction between any two nodes. Scaling exponents in the 
range of $1$-$2$ are obtained through Monte Carlo simulations and 
various clustering coefficients are calculated, one of which,
$C_{\rm out}$, is of order $10^{-1}$,
indicating the network resembles a small-world. The out-degree
distribution has an exponential cut-off for large out-degree.
\end{abstract}


\maketitle

\section{Introduction}

To study the statistical property of a complex system composed of many
interacting individual components, it is often helpful to map the system
into a network of nodes and links (edges). Each node in this network 
represents one component of the real system and the interaction, if there 
is any, between two components is denoted by an edge, 
either directed or undirected, between the two corresponding nodes
in the network.
One quantity of interest is the node-degree profile of the
formed network: How does $n(k)$, the total number of nodes with a 
given number $k$ of links attached (the node degree), scales with $k$? 
Empirical  observations revealed that many social and biological networks
have the scale-free property \cite{barabasi1999,albert2002}, that is,
\begin{equation}
n(k) \sim k^{-\nu}
\label{eq:eq1}
\end{equation}
as $k$ becomes large enough. The scaling exponent $\nu$ is typically
in the range of $2<\nu<3$; but there are evidences that some networks 
have scaling exponents in the range of $1$-$2$
\cite{note1} while a few other networks
have scaling exponents larger than $3$ (for
a collection of experimental data, please refer to Table II of
\cite{albert2002}).

To explain the scale-free characteristic, one appealing mechanism is 
to assume that the network (1) keeps growing and, (2)
during this growth process, new edges are generated and are 
attached preferentially to those nodes which have already been attached
by a large number of edges \cite{barabasi1999}. Based on this mechanism
several models have been suggested 
\cite{barabasi1999,dorogovtsev2000b,krapivsky2000,krapivsky2001}, 
but they predicted the scaling
exponent $\nu$ to be greater than $2$, thus failed to explain the behavior
of those networks with smaller $\nu$. In Ref.~\cite{adamic2000} this 
$``$preferential attachment'' mechanism was questioned partly because
of this apparent discrepancy between theory and empirical data.
In Ref.~\cite{dorogovtsev2001} it was shown that if one assumes a network
is growing accelerately, it is possible to generate scaling exponent
in the range between $3/2$ and $2$. However, it is still not clear 
whether or not this condition is absolutely necessary to explain
experimental observations.

An emerging property of almost all the so-far studied scale-free networks
is that they can at the same time be classified as 
small-world networks \cite{watts1998}. That is,
(i) the diameter of the network scales as $\ln(N)$, where $N$ is the 
total number of nodes in the system, and (ii) the clustering coefficient
$C$ is independent of $N$ and is thus much greater than that of a
random network ($\simeq \langle k \rangle/N$, where $\langle k\rangle$
denotes the average node-degree of the network).
On the theoretical side, it was confirmed that growing networks
generated by the mechanism of preferential attachment will typically
have diameters scales as $\ln(N)$ (see, for example, Ref.~\cite{rodgers2001}).
However, the original Barab\'{a}si-Albert
model \cite{barabasi1999,albert2002} predicted a very small clustering 
coefficient  $C\sim N^{-0.75}$.
The clustering coefficients for other models 
\cite{dorogovtsev2000b,krapivsky2000,krapivsky2001} were not reported.

To improve our understanding on scale-free networks, 
in this work two questions are addressed: Will it be possible to 
generate a scale-free network with scaling exponent  $\nu<2$ based 
on the preferential attachment mechanism?
Will it be possible for a scale-free  network generated this way
to have relatively constant clustering coefficients? 
We answer these questions confirmatively by studying a revised
Krapivsky-Rodgers-Redner model \cite{krapivsky2001} with 
Monte Carlo simulation approach.

\section{The growing network model}

The Krapivsky-Rodgers-Redner model \cite{krapivsky2001} is a generalized 
version of the original Barab\'{a}si-Albert model
on scale-free networks \cite{barabasi1999}.
It has the following key elements: (i)  edges are directed; (ii)
new nodes are added into the network and are attached preferentially
to existing nodes with larger in-degrees; (iii) creation of edges between
$``$old'' nodes are possible and a newly created edge also prefer to
attach to nodes  with larger degrees. 
This model is general in the sense that it takes into account directional
interactions of the real network systems, and that the growth of the
network is not solely caused by the inclusion of new nodes but also
as a result of the increased interactions among the existing nodes of 
the system. It can be corresponded to the real systems including the
World-Wide Web (WWW), the Internet, the food web, the transportation
network, the e-mail network, etc.. Because of its generality, we re-examine 
this model in the present work to illustrate the property of growing
scale-free network.

\begin{figure}[tb]
\psfig{file=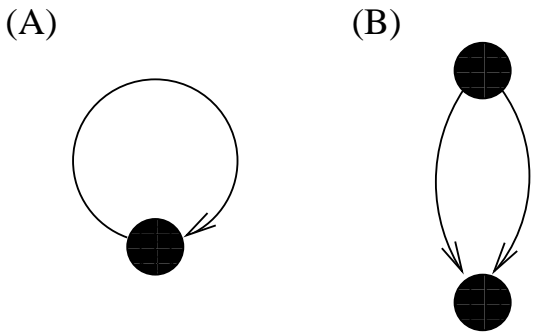,angle=270,width=7.0cm}
\caption{\label{fig:fig01} Self-connection (a) and Multiconnection
of the same direction (b) are forbidden in the present simulation.}
\end{figure}

We noticed in  the original Krapivsky-Rodgers-Redner model
\cite{krapivsky2001} the growth  process permits the following two 
possibilities illustrated in 
Fig.~\ref{fig:fig01}:
(1) a directed edge can originate from and end into the same node
(self-connection) and (2) there can be more than one edges of the
same direction between two nodes (multiconnection). Allowing these
two possibilities makes analytical calculations possible and these
authors found that in the growing network, both the in-degree
and the out-degree distribution follow the power-law \cite{krapivsky2001}:
\begin{equation}
n_{\rm in}(k) \sim k^{-\nu_{\rm in}},\;\;\;\; 2<\nu_{\rm in}=2+p\lambda
<\infty,
\end{equation}
\begin{equation}
n_{\rm out}(k) \sim k^{-\nu_{\rm out}},\;\;
2<\nu_{\rm out}=1+1/(1-p)+\mu p/(1-p)<\infty,
\end{equation}
($p$, $\lambda$ and $\mu$ are the three adjustable model parameters). 
Although the  permission of self-connections and multiconnections might be 
reasonable for some networks (such as WWW pages, in which a page can 
have several links to another page and it can have a link bring the
reader from one portion to  another portion of the same page), it may 
fail for other kinds of 
networks (in a co-authorship network \cite{newman2001}, it is 
meaningless for a author to be the co-author of himself/herself; 
and in a food web \cite{montoya2000}, there is at most one 
edge of the same direction between  two species).
Because of the preferential attachment mechanism, if a node already has
a large number of incoming and outgoing edges, it has 
a good possibility to form self-connections and by doing so
its dominance is further amplified; allowing multiconnection has
similar effects. As a result, edges may concentrate on just few
nodes, making the scaling behavior steeper.
In the present work, the two kind of edges listed in Fig.~\ref{fig:fig01}
are discounted. Because of the reasoning outlined above,
scaling exponents $1<\nu<2$ might occur in a growing network without
self-connection and multiconnection. This point will be checked 
by Monte Carlo simulation.

\section{Monte Carlo Set-up}

The revised Krapivsky-Rodgers-Redner model is studied by MC simulation.
Started with a single node, at each step:
\begin{description}
\item[(i)] With probability $p$, a new node is created and a directed
edge from it to an existing target node $\beta$ is setup.
Of all the $N$ existing  nodes, $\beta$ will be selected 
with probability \cite{krapivsky2001}
\begin{equation}
P_{\rm attach}(\beta)={k_{\rm in}(\beta) +\lambda \over E+\lambda N}.
\label{eq:p_attach}
\end{equation}
In the above equation, $k_{\rm in}(\beta)$, the in-degree, is the total 
number of  incoming edges of node $\beta$; $\lambda$ is a constant
signifying the $``$initial attractiveness'' of a node \cite{dorogovtsev2000b};
and $E$ is the total number of edges in the system before this new 
edge is created.

\item[(ii)] With probability $q=1-p$, a new edge pointing from one
node $\alpha$ to another node $\beta$ is created, provided that
(1) $E<N(N-1)$, (2) $\alpha$ and  $\beta$ are not identical, and 
(3) there is no any preexisting directed edge from $\alpha$ to $\beta$. 
The probability that
$\alpha$ and $\beta$ will be selected is governed by the probability
\begin{equation}
P_{\rm connect}(\alpha,\beta)=
{[k_{\rm out}(\alpha)+\mu][k_{\rm in}(\beta)+\lambda] \over 
\sum\limits_{\gamma}{\sum\limits_{\delta}}^\prime [k_{\rm out}(\gamma)+\mu]
[k_{\rm in}(\delta)+\lambda]},
\label{eq:p_connect}
\end{equation}
where $k_{\rm out}(\alpha)$ denotes the out-degree of node $\alpha$;
$\mu$ is another constant with similar physical meaning as $\lambda$;
${\sum\limits_{\delta}}^\prime$ denotes the summation over all the
nodes $\delta$ which is not yet approached by a directed edge from node 
$\gamma$.
\end{description}

For large system size $N$ it turns out to be quite inefficient
and complicated when performing procedure (ii) based on a direct
application of  Eq.~(\ref{eq:p_connect}). 
This is partly because of the fact that, after a new edge has been created 
between node $\alpha$ and $\beta$, one must update the value of the summation
in Eq.~(\ref{eq:p_connect})  by  $O(N)$ iterations. To speed up procedure 
(ii), the selection of two nodes and the connection of an edge between them is
finished actually  through the following way:
\begin{description}
\item[(1)] Select a outgoing node $\alpha$ with probability
$[k_{\rm out}(\alpha)+\mu]/(E+\mu N)$;
\item[(2)] Select an in-coming node $\beta$ with probability 
$[k_{\rm in}(\beta)+\lambda]/(E+\lambda N)$;
\item[(3)] If $\alpha$ and $\beta$ are identical, or if there is already
a directed edge from $\alpha$ to $\beta$, repeat steps (1) and (2); if
else, accept $\alpha$ and $\beta$ and update the system.
\end{description}
It is not difficult to prove that by this method the  probability for
nodes $\alpha$ and $\beta$ be chosen is identical to  
Eq.~(\ref{eq:p_connect}).

The algorithm code is written in C++ language \cite{stroustrup1997}, 
with some of its standard containers (including {\it map} and {\it set}) 
being exploited.

To estimate the scaling exponents from the simulated data, we  use
two methods. One can
directly fit the data with Eq.~\ref{eq:eq1}. Alternatively, one can
define the cumulative degree distribution by
\begin{equation}
P(k)=\sum\limits_{k^\prime\geq k} n(k^\prime) \sim k^{-(\nu -1)},
\label{eq:eq2}
\end{equation}
and from the cumulative distribution data an estimation of the
value of $\nu$ could be obtained.

\section{ \label{sec:degree} Degree distribution of the growing network}

\begin{figure}[tb]
\psfig{file=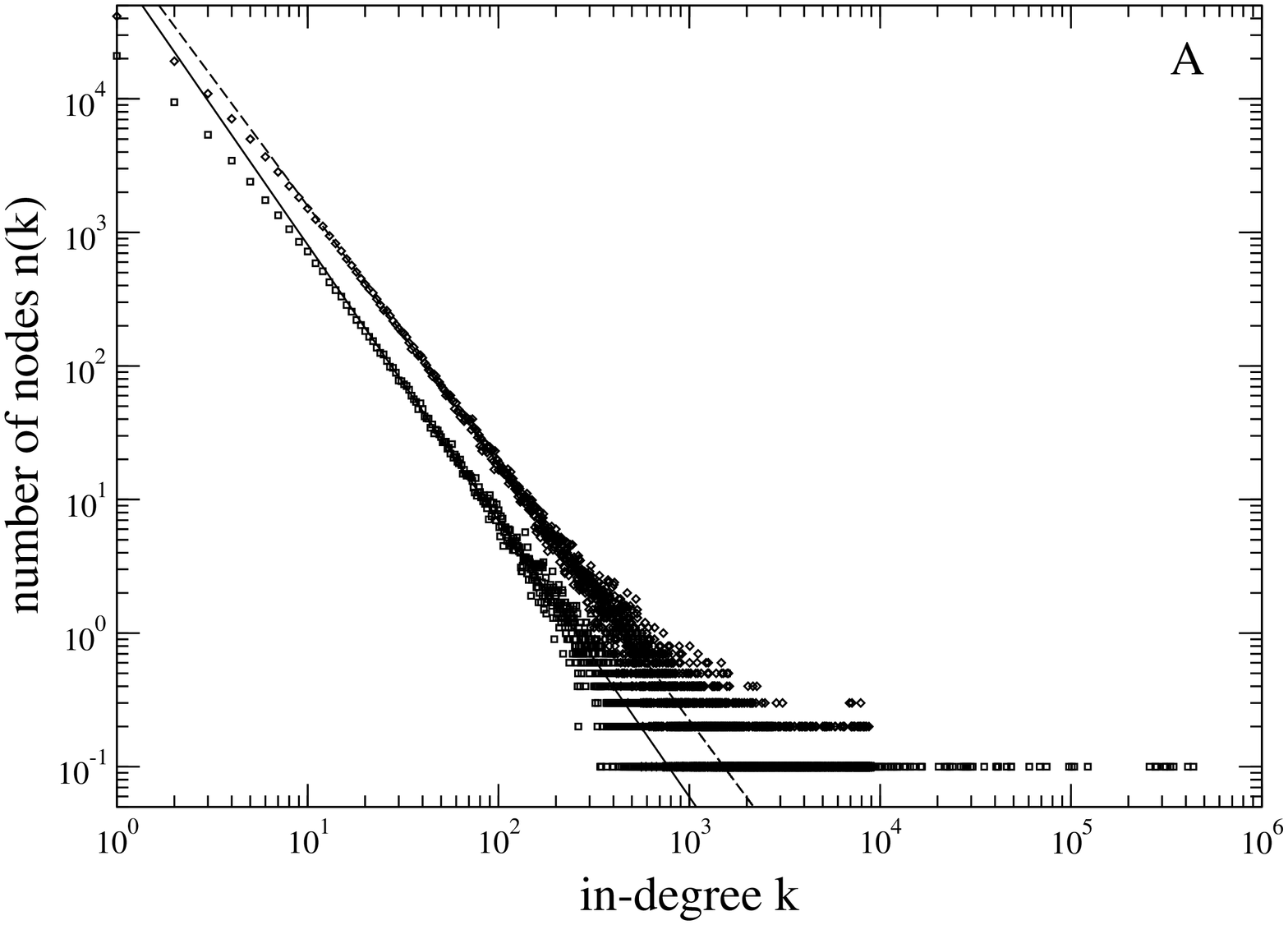,angle=270,width=8.0cm}

\psfig{file=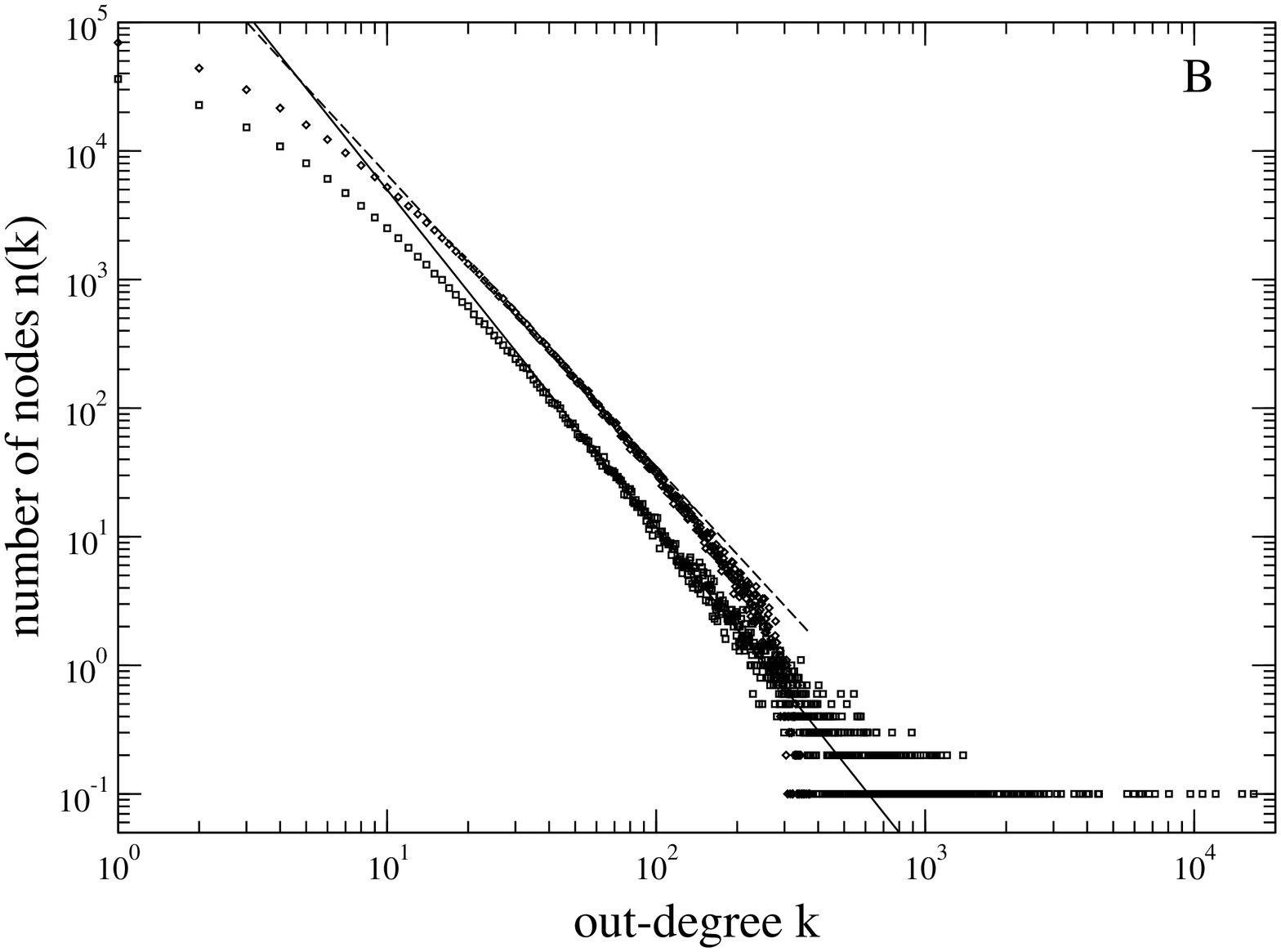,angle=270,width=8.0cm}
\caption{\label{fig:fig02} The profiles of in-degree (A) and
out-degree (B) distribution at $p=0.133334$, $\lambda=0.75$ and
$\mu=3.55$, after a growing process of $10^6$ steps.
Square symbols are the data for 
the original Krapivsky-Rodgers-Redner model and diamonds are
the data for the revised model. Each data point is the average over
$20$ (diamonds) or $10$ (squares) realization of the network.
The thin solid line has a slop of $-2.066$ in
(A) and $-2.626$ in (B). The thin dashed line
has a slop of $-1.925$ in (A) and $-2.269$ in (B).
The average number of nodes in the revised network is $133271$, 
and the average number of edges is $999984$.
}
\end{figure}

\begin{figure}[tb]
\psfig{file=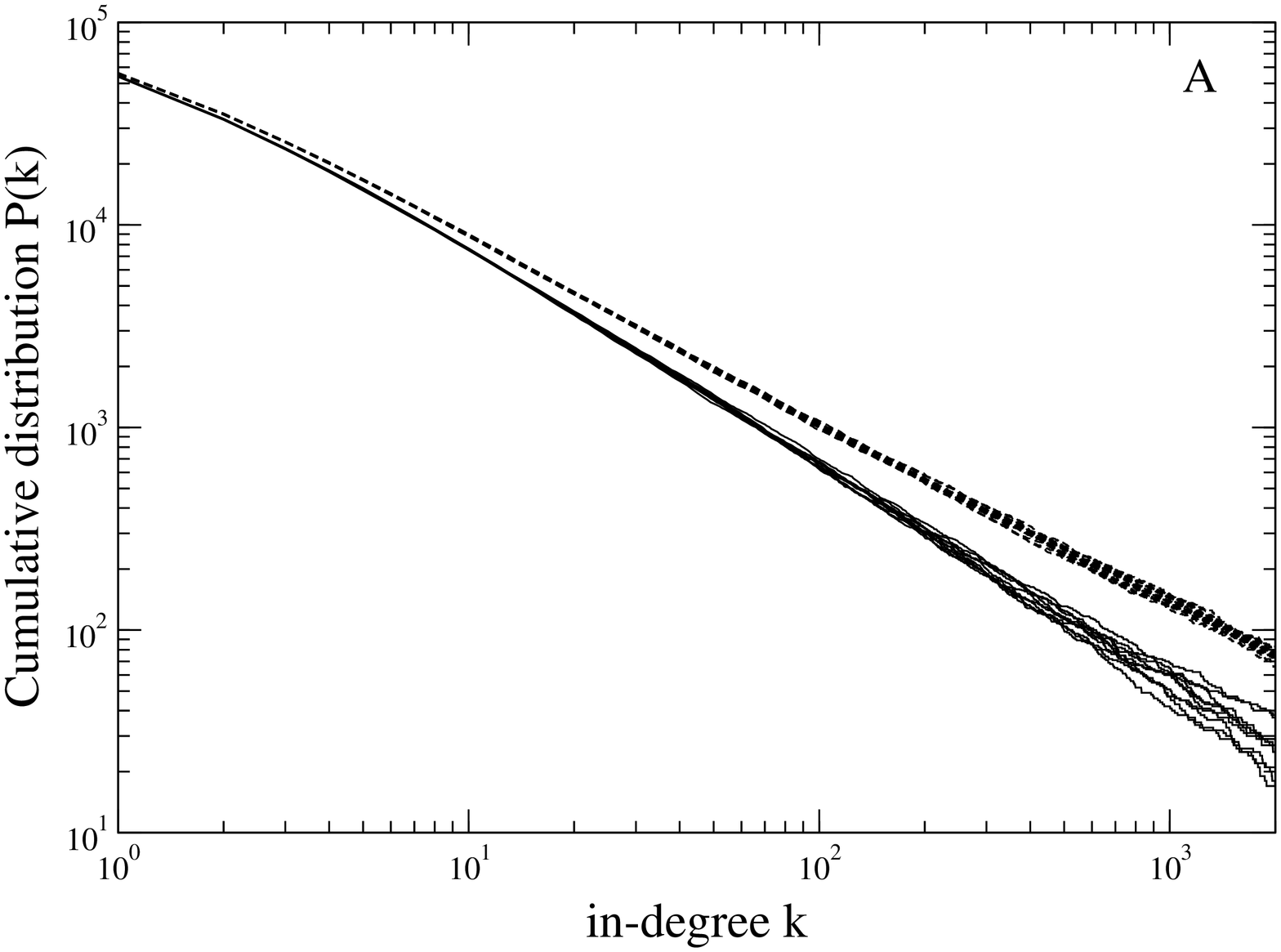,angle=270,width=8.0cm}

\psfig{file=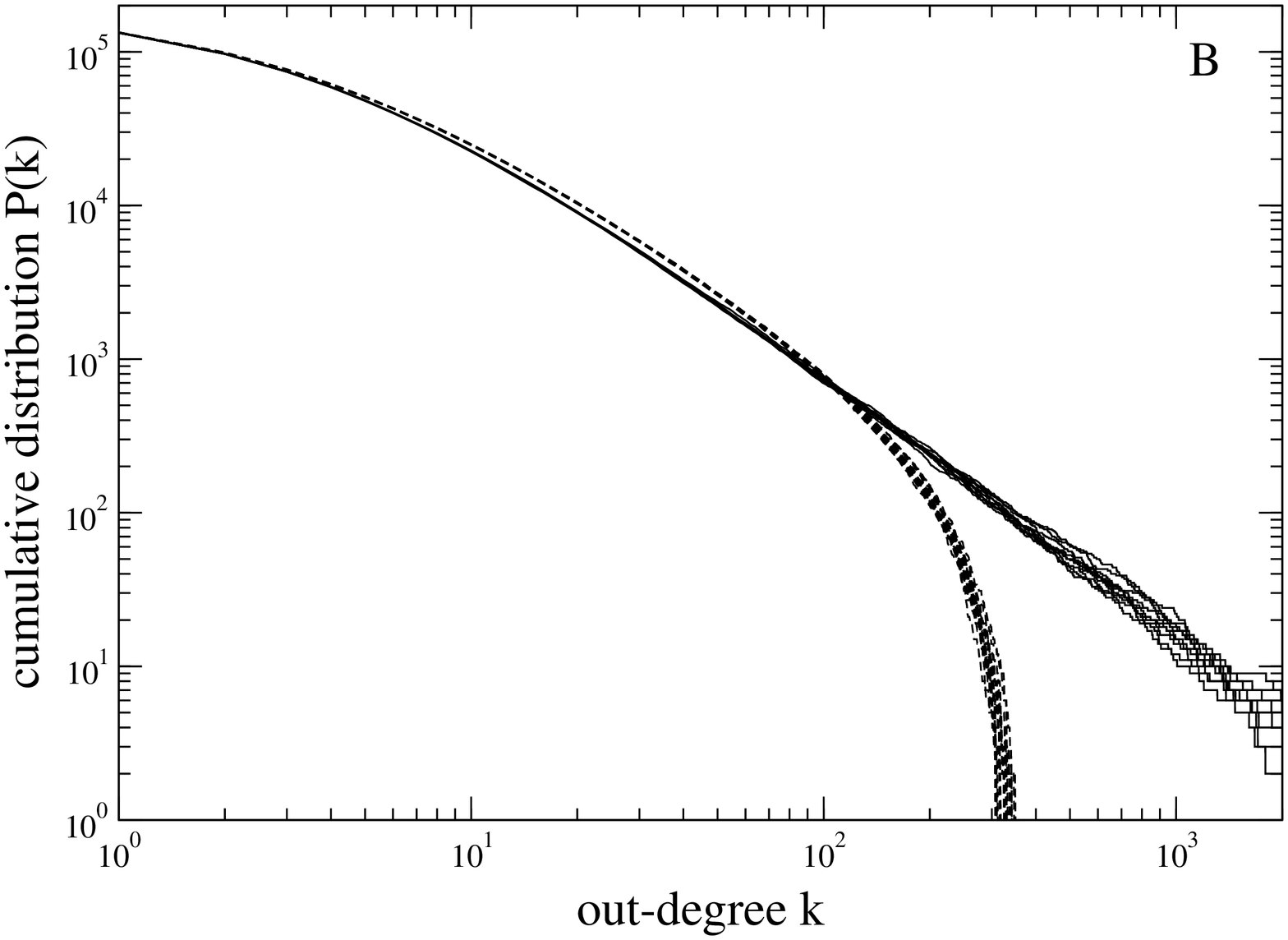,angle=270,width=8.0cm}
\caption{\label{fig:fig03} In-degree (A) and out-degree (B)
cumulative distribution
for the data sets in Fig.~\ref{fig:fig02}. The solid lines 
corresponds to the original model, and the dashed lines to the
revised model. From these curves, we estimate 
$\nu_{\rm in}=2.066\pm 0.014$ (the original model) and
$\nu_{\rm in}=1.925\pm 0.007$ (the revised model).
$\nu_{\rm out}=2.626\pm 0.036$ for the original model.
The out-degree cumulative distribution of the revised model does not fit
well to the power-law. Therefore, the out-degree scaling
exponent is not estimated by this method but by a direct fit
to the distribution in Fig.~\ref{fig:fig02}B, resulting
in $\nu_{\rm out}=2.269$. 
}
\end{figure}

In the growing network model, there are three adjustable parameters,
namely $p$, $\lambda$, and $\mu$.  Figure \ref{fig:fig02}
shows  the relations between the  average number of nodes  and 
the in-degree and out-degree for both the original and the revised
Krapivsky-Rodgers-Redner model. Figure \ref{fig:fig03} shows the
corresponding cumulative degree distributions for the two models.
The network is the result of $N=10^6$ growing steps.

At $p=0.133334$ (a new node
will be included on average every $7.5$ steps), $\lambda=0.75$
and $\mu=3.55$ the original model \cite{krapivsky2001} predicts 
$\nu_{\rm in}=2.1$ and $\nu_{\rm out}=2.7$.
From the MC data we obtain that 
$\nu_{\rm in}=2.066\pm 0.014$ and $\nu_{\rm out}=2.626\pm 0.036$,
in close agreement with the analytical values.
At these same parameters, the revised model has
$\nu_{\rm in}=1.925\pm 0.007$ and $\nu_{\rm out}=2.269$.
Thus, excluding of self-connection and multiconnection leads to 
decreased values for the scaling exponents. 
Other quantitative differences are: 
\begin{description}
\item[(1)]
In the revised model there is a cut-off in the in-degree distribution:
no node has in-degree $k>9\times 10^3$.  While in the
original model, there are nodes with in-degree as large as $k=4\times 10^5$.

\item[(2)]
In the revised model the out-degree distribution has an exponential cut-off
around $k=250$; while in the original model, there are nodes with out-degree
as large as $k=1.5*10^4$.

\item[(3)]
The value of $n(k)$ is much larger in the revised model than in the
original model for a given $k$ (less than the cut-off value). This
holds both for the in-degree distribution and for the out-degree
distribution. 
\end{description}
These observations lead to the following picture:
By prohibiting self-connection and multiconnection, edges which 
originally belong to just few $``$suppernodes'' are now redistributed
to those nodes of small or intermediate in- and out-degrees. 
Consequently, the number of nodes with small and intermediate
node-degrees increases considerably, resulting in a smaller
scaling exponent in the power-law decrease of the distribution
and a cut-off in the tail of this distribution.

\begin{figure}[tb]
\psfig{file=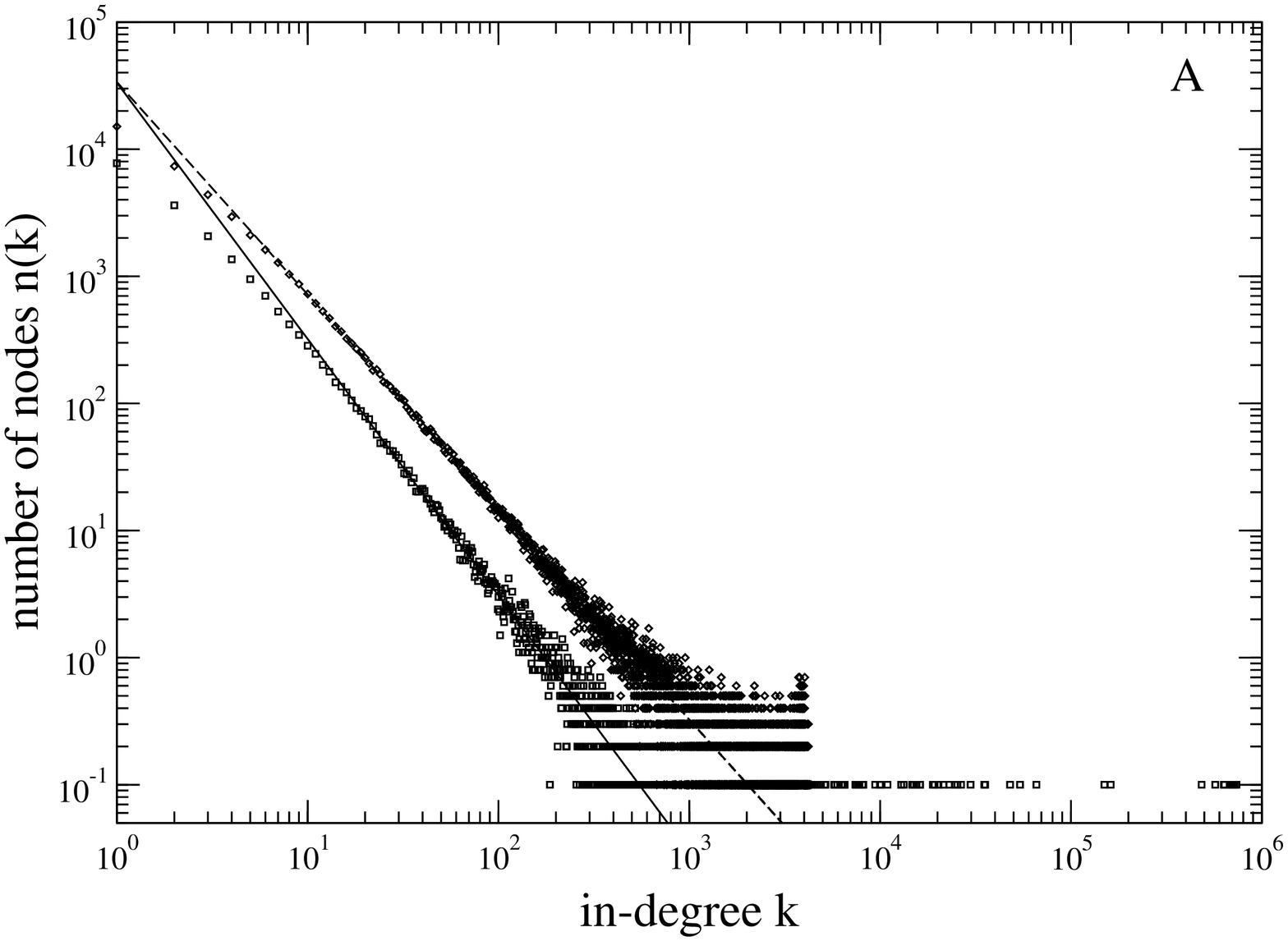,angle=270,width=8.0cm}
\psfig{file=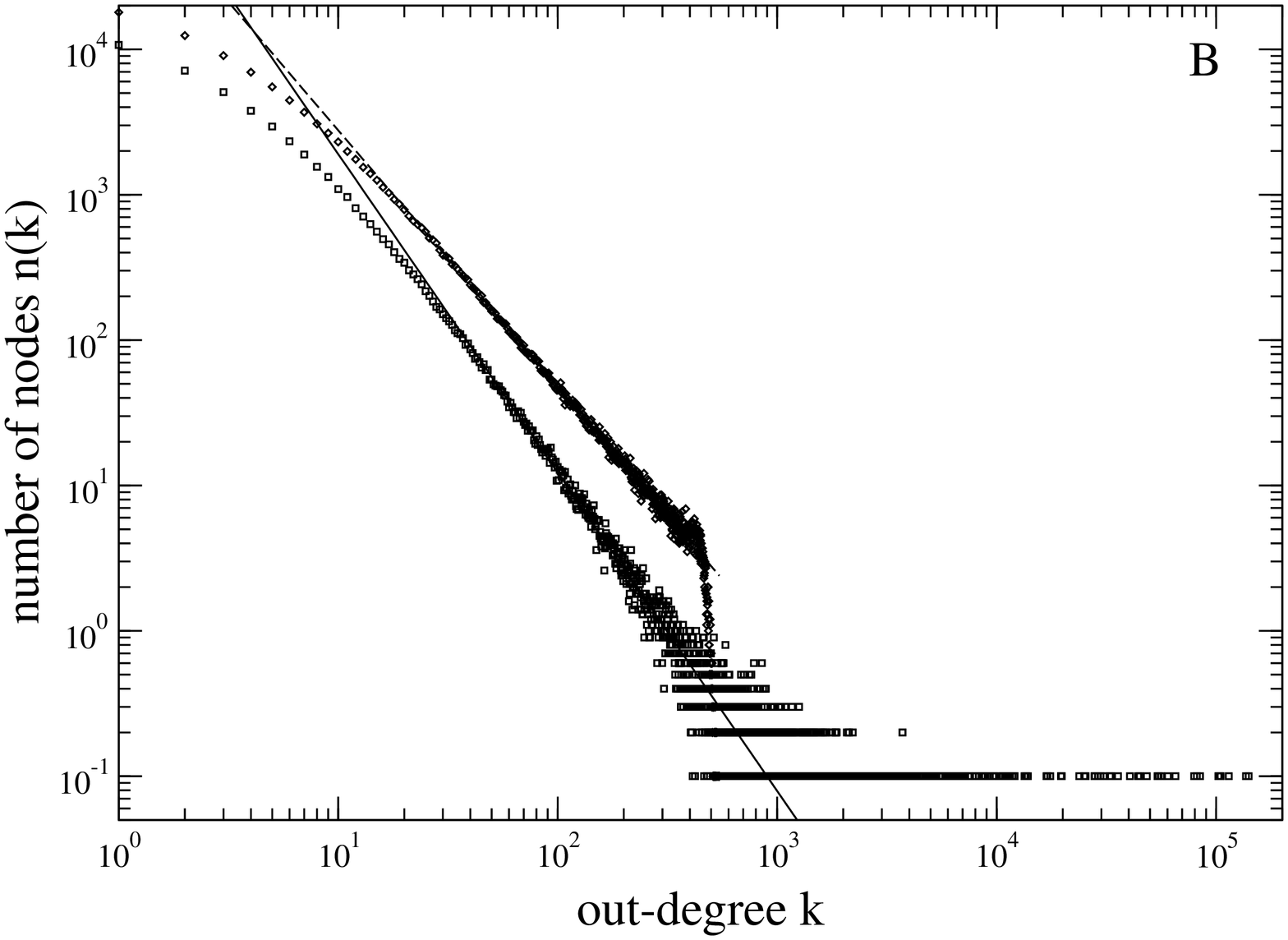,angle=270,width=8.0cm}
\caption{\label{fig:fig04} The profiles of in-degree (A) and
out-degree (B) distribution at $p=0.05$, $\lambda=0.75$ and
$\mu=3.55$ and $10^6$ growing steps. 
Squares (averaged over $10$ realizations) correspond to the
original model and diamonds (averaged over $20$ realizations)
to the revised model. The thin solid line has slope
$-2.018$ in (A) and $-2.190$ in (B). The thin dashed line has slope
$-1.672$ in (A) and $-1.764$ in (B).
The average number of nodes
is $49831$ and the average number of edges is $999860$.}
\end{figure}

In Fig.~\ref{fig:fig04} we demonstrate the simulation result when the
probability of node addition is changed to $p=0.05$ (a new node
will be included on average every $20$ steps) while the other two parameters
are kept the same values as in Fig.~\ref{fig:fig02}. In these parameters,
the original model predicts $\nu_{\rm in}=2.04$ (theory) and
$2.018\pm 0.015$ (MC) and $\nu_{\rm out}=2.24$
(theory) and $2.190\pm 0.014$ (MC); while the revised model 
has an in-degree exponent $\nu_{\rm in}=1.672\pm 0.003$ and an 
out-degree exponent $\nu_{\rm out}=1.764$,  both of which are markedly 
smaller than $2$. 

Therefore, the exclusion of self-connection and multi-connection can
change the scaling exponent of the scale-free network dramatically when 
each node has a relatively large average node-degrees. It can be 
anticipated that similar behavior will be observed when the initial
attractiveness parameters, $\lambda$ and $\mu$, of each node are 
varied.  It is therefore possible for the present model to explain
networks with scaling exponent $\nu<2$.

\begin{figure}
\psfig{file=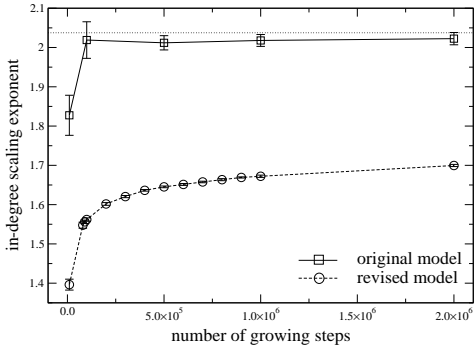,angle=270,width=8.0cm}
\caption{\label{fig:fig05} The relation between the in-degree
scaling exponent and growing steps for the original (squares) and the
revised (circles) model. The parameters are set equal to that
of Fig.~\ref{fig:fig04}, namely, $p=0.05$, $\lambda=0.75$ and
$\mu=3.55$. The thin dotted line indicates the theoretical
prediction of $\nu_{\rm in}=2.075$ for the original model.}
\end{figure}

However, could the observation of scaling exponents $1<\nu<2$ be 
an artifact caused by finite-size effects?  In Fig.~\ref{fig:fig05}
the calculated in-degree scaling exponent is plotted as a
function of the total growth steps. This figure strongly indicates
that even for an infinite system the scaling exponents will
still be less than $2$.

Adamic and Huberman \cite{adamic2000} studied the WWW by
mapping each web domain (rather than each web-page) as a single node,
and they reported an in-degree scaling exponent of
$\nu_{\rm in}=1.94$. 
Mossa and co-authors, upon their reinterpretation of the WWW data of
Barab\'{a}si {\it et al.} \cite{barabasi1999}, reported an exponent of
$\nu=1.25$ \cite{mossa2002}. The e-mail network studied in
Ref.~\cite{ebel2002} has a scaling exponent in the range $1.47<\nu<1.82$.
And even smaller scaling exponents are
reported in several other networks \cite{albert2002}. To 
attain quantitative agreement to these empirical data, one need nevertheless
more information to help fixing the values of the adjustable parameters.

A persistent property of the revised network model is that there is
a rapid decay in the out-degree distribution (which occurs at $k_{\rm out}
\simeq 400$ in Fig.~\ref{fig:fig04}B). Such a rapid decay was not
observed in the original model.
This feature is also absent in the in-degree distribution of both the
revised and the original model, although the in-degree distribution
of the revised model does have a cut-off for large $k$. 
Experimentally, it was reported that both
the WWW network \cite{huberman1999,mossa2002} and the e-mail network
\cite{ebel2002} show exponential cut-off in the node-degree distribution.

\section{\label{sec:clique} Cluster coefficients of the growing  network}

As was mentioned in the introduction, many real scale-free network 
at the same time show small-world behavior \cite{albert2002}: 
having small diameters and being highly  clustered.
For the original Krapivsky-Rodgers-Redner model, it has been reported 
in Ref.~\cite{rodgers2001} that the average minimum path scales as 
$\ln(N)$. We anticipate
this to be  hold also for the revised model. Here we
focus on the clustering characteristics of the revised model system, with
the parameters setting equal to those of Fig.~\ref{fig:fig02}, i.e.,
$p=0.133334$, $\lambda=0.75$ and $\mu=3.55$, and a total $10^6$ 
growing steps.

\subsection{Mutual-connection coefficient $C_{\rm mutual}$}

Denote $G_{\rm down}(\alpha)$ as the {\it complete}  set of nodes 
which are the $``$downstream'' neighbors of node $\alpha$, namely 
there exists a  directed edge from $\alpha$ to each nodes 
in $G_{\rm down}(\alpha)$;
similarly we define $G_{\rm up}(\alpha)$ as the complete set of 
nodes which are the $``$upstream'' neighbors of node $\alpha$.
$|G_{\rm down}(\alpha)|$ means the size of set $G_{\rm down}(\alpha)$.

The mutual-connection coefficient is defined as 
\begin{equation}
C_{\rm mutual}={\sum_{\alpha} 
|G_{\rm down}(\alpha)\bigcap G_{\rm up}(\alpha)|/|G_{\rm down}(\alpha)|
\over N},
\end{equation}
which signifies to what extent the $``$downstream neighbors'' of one
node intersect with its $``$upstream neighbors''.  
The value of $C_{\rm mutual}$ averaged
over $20$ realization of the growing network is $0.0010$. This indicates
that the interaction between one node and its $``$neighbors''
in the network is usually not bi-directional.  However, this value is 
still  much larger than the
value for a random network of the same size ($N=133271$) and  the same 
average degree of out-going edges ($\langle k\rangle =7.50$), 
for which $C_{\rm mutual}=5.63\times 10^{-5}$.

\subsection{Incoming clustering coefficient $C_{\rm in}$}

Suppose a given node $\alpha$ has in-degree $k_{\rm in}(\alpha)$.
The maximal number of edges existing between the nodes in $G_{\rm up}(\alpha)$
is $k_{\rm in}(\alpha)(k_{\rm in}(\alpha)-1)$.
Denote $i_{\rm actual}(\alpha)$ as the actual number of 
edges existing
between these edges. We define the 
incoming clustering coefficient as
\begin{equation}
C_{\rm in}={{\sum\limits_{\alpha}}^\prime
i_{\rm actual}(\alpha)/[k_{\rm in}(\alpha)(k_{\rm in}(\alpha)-1)]
\over N^{\prime}}
\end{equation}
where ${\sum\limits_{\alpha}}^\prime$ indicates summation over
all the nodes whose incoming edges is larger than $1$, and 
$N^{\prime}$ is the total number of nodes with this property.
We find that $C_{\rm in}=0.0044$. This value indicates that the degree of
cliqueness of the upstream neighbors of a 
given node is usually very small. For a completely random
graph, $C_{\rm in}=5.63\times 10^{-5}$.

\subsection{\label{sub:clique_out} Outgoing clustering coefficient
$C_{\rm out}$}

The definition of the out-going clustering coefficient 
$C_{\rm out}$ is similar to that of $C_{\rm in}$.
Suppose a particular node $\alpha$ has $k_{\rm out}(\alpha)$  out-going
edges, and $i_{\rm actual}(\alpha)$ is the total number of edges
between the nodes in  $G_{\rm down}(\alpha)$, then 
\begin{equation}
C_{\rm out}={{\sum\limits_{\alpha}}^\prime 
i_{\rm actual}(\alpha)/[k_{\rm out}(\alpha)(k_{\rm out}(\alpha)-1)]
\over N^{\prime\prime}},
\end{equation}
where $N^{\prime\prime}$ is the total number of nodes whose out-degree is
larger than $1$. 
We find $C_{\rm out}=0.229$ for the present growing network. 
Compared with the small values of $C_{\rm mutual}$ and $C_{\rm in}$, such
a large value of $C_{\rm out}$ is surprising. It suggests the average
interaction between the nodes which are in the
downstream group of a given node is considerably strong. 
How to  understand this kind of asymmetry, namely
$1\sim C_{\rm out}\gg C_{\rm in}$?  We suggest the following possibility:
In the network, there are some nodes which are so popular that
a large population of the whole nodes  will have an edge pointing to them
(see Fig.~\ref{fig:fig02}A).
Consequently, these nodes will have great possibility to belong to the
downstream group of any particular node, and they will also
have great possibility to be pointed to by other members of this group,
making $C_{\rm out}$ to be proportional to unity. 
However, the number of nodes decays quickly when the out-degree increases
to about $250$ (see Fig.~\ref{fig:fig02}B). Therefore, in the network
there is no node which are so ``generous'' that it points to a large
population of the whole network. This may make the value of $C_{\rm in}$
small. In other words, it might be the existence of an steep cut-off in
the out-degree profile that accounts for the difference in the 
clustering coefficients $C_{\rm out}$ and $C_{\rm in}$.

\subsection{Triangle coefficient $C_{\rm triangle}$}

For a particular node $\alpha$, suppose node $\beta \in 
G_{\rm down}(\alpha)$. Then 
$i_{\triangle}(\alpha,\beta)=|G_{\rm down}(\alpha) 
\bigcap G_{\rm down}(\beta)|$
is the total number of nodes that are pointed to by both $\alpha$ and
$\beta$. We define
\begin{equation}
C_{\rm triangle}={\sum\limits_{\alpha}
\sum\limits_{\beta\in G_{\rm down}(\alpha)}
i_{\triangle}(\alpha,\beta)/k_{\rm out}(\beta) \over N}.
\end{equation}
The triangle coefficient $C_{\rm triangle}$ signifies to what extent,
if there is a directed edge from node $\alpha$ to node $\beta$ and there 
is a directed edge from $\beta$ to node $\gamma$, there
will also be a directed node from $\alpha$ to $\gamma$.
Calculation revealed that $C_{\rm triangle}\simeq 0.011$.

\section{\label{sec:discussion} Conclusion and  Discussion}

In this work we have used a revised Krapivsky-Rodgers-Redner model
to investigate the degree distribution of growing random network and
to investigate whether such a kind of growing  network could be
regarded as a small-world network. After excluding the possibility of 
self-connection (Fig.~\ref{fig:fig01}A) and  requiring that there
is at most one directed edge from one node to any another node 
(Fig.~\ref{fig:fig01}B),
the Monte Carlo simulation demonstrated that scale-free network with
degree distribution coefficient $\nu$ less than $2$ can be generated.
And it is also revealed that the average interactions between the nodes
which are belong to the downstream group of a particular node is 
very strong, suggesting the growing network at the same time
forms a $``$small world''. The strong interaction in the downstream group of a 
particular node was suggested to be closely related to the existence
of several $``$popular'' nodes which are pointed to by a large fraction
of the total population in the node system.  Previous efforts often
predicted that the scaling exponent $\nu$ should be larger than $2$,
and there is still not many efforts to understand why many scale-free
networks are at the same time small-world networks. It is hoped that
the present work will help to improve our understanding of the occur of
scale-free networks with $\nu<2$ and to improve our understanding
of the close relationship between scale-free and small-world networks.

The present work suggests that, by excluding the possibility of
self-connection and multiconnection, those many edges, which were
associated with several nodes of extremely large in- or out-degrees
in the original Krapivsky-Rodgers-Redner network, are now 
redistributed to nodes of small or intermediate degrees. This
may explain why a dramatic decrease in scaling exponents could
be observed.  

\section{Acknowledgement}

This work was financially supported by the Alexander von Humboldt
foundation. The author is grateful for the hospitality of Professor
Reinhard Lipowsky.

\newpage

\begin{thebibliography}{16}
\bibitem{note1}
If the node degree distribution looks like a power law with
the exponent smaller than $2$, its tail must have a different
exponent or have some sort of cut-off, since the average number
of edges connected to one node should be finite. A cut-off in
the distribution is observed in the present work.
\end{thebibliography}

\end{document}